\documentclass[
aps,
prb,
twocolumn,
groupedaddress
]
{revtex4-1}
\usepackage{siunitx}
\usepackage{graphicx}
\usepackage{amsmath, amssymb}

\newcommand{\tfg}[1]{\textsubscript{\protect\raisebox{-1pt}{#1}}}

\newcommand{\mtfg}[1]{_{\mathrm{#1}}}			
\newcommand{\mhog}[1]{^{\mathrm{#1}}}

\newcommand{\mhtg}[2]{\mhog{#2}\mtfg{#1}}
\newcommand{\diff}{\mathrm{d}}

\begin{document}
	
	
	\title{Polaronic inter-acceptor hopping transport in intrinsically doped nickel oxide}
	
	
	\author{Robert Karsthof}
	\email[]{robert.karsthof@physik.uni-leipzig.de}
	\affiliation{Felix Bloch Institute for Solid State Physics, Universit\"{a}t Leipzig, Linn\'{e}str. 5, 04103 Leipzig, Germany}
	
	\author{Arthur Markus Anton}
	\affiliation{Peter Debye Institute for Soft Matter Physics, Universit\"{a}t Leipzig, Linn\'{e}str. 5, 04103 Leipzig, Germany}
	
	\author{Friedrich Kremer}
	\affiliation{Peter Debye Institute for Soft Matter Physics, Universit\"{a}t Leipzig, Linn\'{e}str. 5, 04103 Leipzig, Germany}
	
	\author{Marius Grundmann}
	\affiliation{Felix Bloch Institute for Solid State Physics, Universit\"{a}t Leipzig, Linn\'{e}str. 5, 04103 Leipzig, Germany}

	
	\date{\today}
	
	\begin{abstract}
		In this work, we revisit the issue of the nature of electronic transport in nickel oxide (NiO) and show that the widely used model of free small polaron hopping, initially raised to characterize transport in high-purity samples, is not appropriate for modeling intrinsically doped NiO. Instead, we present extensive evidence, collected by means of temperature- and frequency-dependent measurements of the electrical conductivity $\sigma$, that the model of polaronic inter-acceptor hopping can be used to consistently explain the electronic conduction process. In this framework, holes are localized to acceptors (Ni vacancies), forming a strongly bound, polaron-like state. They can only move through the film by hopping to a neighboring, at least partially unoccupied, acceptor. This renders the spatial overlap between neighboring polaronic wave functions a highly critical parameter. The signature of this process is the occurrence of two temperature regions of the DC conductivity, separated by about half the Debye temperature $\frac{\theta\mtfg{D}}{2} \approx \SI{200}{\kelvin}$. For $T>\frac{\theta\mtfg{D}}{2}$, holes are transferred by phonon-assisted hopping over the potential barrier between two sites, whereas phonon-assisted tunneling through the barrier dominates below that temperature. We also show that the degree of structural and electronic disorder plays a vital role in determining the characteristics of the transport process: high disorder leads to strong energetic broadening of the acceptor states such that hopping to more distant sites may be favored over transfer to nearest neighbors (variable range hopping).
		The assumption of high binding energies of the charge carriers at V\tfg{Ni} is in accordance with the recent paradigm shift regarding the understanding of the electronic structure of NiO: holes doped into NiO couple to Ni $3d$ spins, thereby occupying deep polaron-like states within the band gap (Zhang-Rice bound doublets). This work is the first to explicitly take this perception into account to explain carrier transport in NiO.

	\end{abstract}
	
	
	\maketitle
	
	\section{Introduction}
	
	The long-standing question on the origin of the DC conduction mechanism in nickel oxide has never been entirely settled. There exist numerous studies regarding this subject, many of which support the picture of polaronic conduction \cite{Wagner.1933,Boer.1937,Heikes.1957,Houten.1962,Austin.1967,Bosman.1970,Adler.1970}, i.e. the hole is surrounded by a self-induced lattice distortion. Depending on the size of this distortion, the charge carrier can move by thermally activated hopping (for small polarons) or by conduction in a narrow band (mostly for large polarons when spatial overlap of the wave functions is appreciable). Evidence for both models have been presented in the literature. Austin and Mott have conjectured that the true value for the spatial extent $r\mtfg{p}$ of the polaronic wave function lies somewhere between the 'small' and 'large' case \cite{Austin.1967}. This view has also been taken by Adler and Feinleib \cite{Adler.1970}, based on an estimate of $r\mtfg{p}$ in the order of two lattice spacings. On the other hand, AC studies have shown that \textit{bound} charge carriers have probably small polaron character \cite{Snowden.1964,Kabashima.1968,Aiken.1968,Kolber.1972}. 
	An entirely different picture has been proposed by Lunkenheimer \textit{et al.} who have interpreted their AC conductivity data within the framework of the \textit{correlated barrier hopping} model \cite{Lunkenheimer.1991}. In this model, the charge carriers are thought to hop from one acceptor site to another, with the hopping energy critically depending on the acceptor density \cite{Pike.1972}, resembling the so-called impurity conduction in the hopping regime. In fact, there are several scientific publications from the 1960s conjecturing about the DC transport mechanism in NiO being due to inter-acceptor (or impurity) hopping conduction \cite{Nachman.1965,Springthorpe.1965,Austin.1967,Aiken.1968,Kabashima.1968,Bosman.1970}. However, this view seems to have been abandoned later on in favor of "free" small or large polarons that can be transferred by hopping to any (sub-)lattice site. To our view, this was mainly the result of an increased availability of low-impurity NiO samples at that time with considerably higher charge carrier mobilities than before. This demanded for an alternative conduction model. In the recent decades, researchers have aimed to use NiO in different (opto-)electronic devices which, in many cases, requires the material to be at least moderately conductive. To this end, doping (intrinsically or extrinsically) is often a prerequisite which should bring the impurity (or more general: acceptor) hopping conduction model into focus again.
	
	The perception of the hole in NiO was fundamentally changed in 1994, when Ba{\l}a, Ole\'{s} and Zaanen \cite{Bala.1994} applied a theory to NiO that was originally developed for high-$T\mtfg{c}$ superconductors by Zhang and Rice \cite{Zhang.1988}. It considers the coupling of the spin of a hole (introduced either by doping or by excitation) to the majority $3d$ spins at a neighboring Ni site. This coupling, which was shown to be of antiferromagnetic nature, leads to the formation of a bound state, the so-called Zhang-Rice (ZR) bound state, with total spin $S=\frac{1}{2}$ in the case of NiO and thus being a doublet. This new view made it possible to explain as yet inexplicable features in photoelectron spectroscopy experiments, where the ZR doublet appears as the state with the lowest ionization energy. Taguchi \textit{et al.} \cite{Taguchi.2008} also predicted an important role of this state in the electrical conduction process in NiO. However, to our knowledge, no attempt has been made so far to develop a conduction model taking the properties of ZR states explicitly into account. The high binding energies of the holes occupying these states imply that carrier transport takes place exclusively through these energetically rather isolated states without the participation of any other states or bands. Thus, the compatibility of the ZR with the inter-acceptor hopping model is given, and the validation of this attempt is the aim of the current work. \\
		
	Recently, NiO has re-emerged as a material of special interest in the device community. It has found application in organic solar cells \cite{He.1999,Irwin.2008,Park.2010}, light-emitting diodes \cite{Park.2005,Tang.2013}, resistive-switching devices \cite{Seo.2005,Kim.2008}, perovskite solar cells \cite{Jeng.2014} and electrochromic devices \cite{Avendano.2006,Huang.2011,Moulki.2012,Ren.2013}. In many applications, efficient hole transport through the NiO layer is essential while, at the same time, the layer itself should be transparent in the visible and near-infrared spectral range. It is well known that the conductivity of NiO can be increased by intrinsic doping with Ni vacancies (through high oxygen supply during film growth) or by extrinsic doping, e.g. with group-I elements like Li \cite{Lany.2007,Park.2010}. However, as the doping level is increased, the optical transmittance of NiO typically shows a marked decrease \cite{Newman.1959} which has recently been attributed to the development of a broad ZR band deep inside the band gap \cite{Zhang.2018}. This behavior requires a trade-off between high optical transparency and electrical conductivity. The latter is typically not higher than \SI{0.1}{\siemens\per\centi\meter} due to low hole mobilities ($\mu\mtfg{p} \le \SI{0.1}{\square\centi\meter\per\volt\per\second}$) of the localized charge carriers. In order to design NiO layers addressing the requirements for simultaneous conductivity and transparency, an understanding of the microscopic nature of carrier transport processes is necessary.

	\section{Experimental methods}
	 
	The NiO films were prepared by two different methods. Pulsed laser deposition (PLD) growth was carried out in an in-house-built system, using a KrF excimer laser (wavelength \SI{248}{\nano\meter}, pulse energy \SI{650}{\milli\joule}) ablating a ceramic NiO target (purity \SI{99.998}{\percent}, Alfa Aesar). PLD growth was done either at room temperature (no intentional substrate heating) or at a substrate temperature of \SI{300}{\degreeCelsius}. In both cases, the partial pressure of the oxygen background atmosphere was set to \SI{0.1}{\milli\bar}. Reactive DC magnetron sputtering from a metallic Ni target in an Ar/O\tfg{2} atmosphere was additionally employed to grow NiO films. Layer thicknesses between \SIlist{100;350}{\nano\meter} have been achieved and checked by means of profilometry. $c$-plane-oriented, one-sided polished Al\tfg{2}O\tfg{3} single crystals were used as substrates. Metallic back contact layers consisting of Pt (thickness $\approx \SI{50}{\nano\meter}$) were deposited by DC magnetron sputtering in Ar atmosphere prior to NiO growth. The NiO layers were directly capped by a \SI{20}{\nano\meter} thick Pt layer, and both NiO and Pt capping were patterned into pillar-shaped contacts (diameters between \SI{250}{\micro\meter} and \SI{800}{\micro\meter}) by standard UV photo-lithography and the lift-off technique.
	
	Current-voltage characterization of the Pt/NiO/Pt structures was carried out with the help of a SÜSS WaferProber with tungsten needles and an AGILENT 4155C Precision Semiconductor Parameter Analyzer. On the basis of these measurements, individual contacts were selected for further characterization. Those samples were then mounted onto transistor sockets, and the selected pillars were contacted by Au wire bonding using silver epoxy resin. Temperature-dependent current-voltage measurements on thusly prepared samples were done in a closed-cycle He cryostat.

	Broadband dielectric spectroscopy (BDS) measurements have been accomplished on the same samples as used for current-voltage studies. The spectra have been recorded in a temperature and frequency range of \SIrange{120}{300}{\kelvin} and \SIrange{e-2}{e7}{\hertz}, respectively, employing a NOVOCONTROL Technology high-resolution $\alpha$-analyzer combined with a Quatro temperature controller ensuring absolute thermal stability of $\le \SI{1}{\kelvin}$. 
	
	\section{Results}
	
	\subsection{Structural properties}
	
\begin{figure}
		\centering
		\includegraphics[width=\columnwidth]{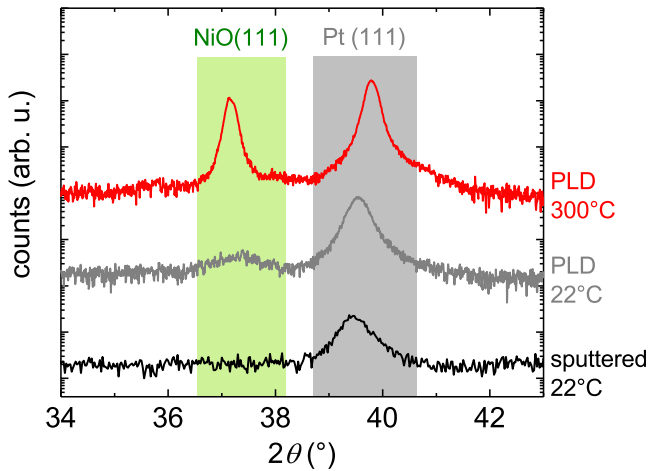}
		\caption{X-ray diffraction pattern of the (111) reflexes of NiO grown on Pt/fused silica for PLD-grown and sputtered thin films at room temperature, as well as PLD-grown at elevated temperature.}
		\label{fig:XRD_PLD+ms_comp}
	\end{figure}

	 In the past, NiO has been deposited at room temperature in our laboratories using both methods, either PLD or sputtering, in order to fabricate active electronic devices \cite{Karsthof.2015,Karsthof.2015a,Karsthof.2016}. However, these films exhibit high structural disorder, as evident from $2\theta$-$\omega$ X-ray diffraction scans shown in Fig.~\ref{fig:XRD_PLD+ms_comp}. While room temperature-PLD on top of a Pt back contact leads to nano-crystalline films where the (111) reflex of NiO can be observed, this peak is absent for the sputtered films. Using the Scherrer formula
	
	\begin{equation}
	l\mtfg{z} \approx \frac{0.9\lambda}{\Delta(2\theta)\cos(\theta)}
	\label{eq:scherrer}
	\end{equation}
	
	with $\lambda = \SI{5.406}{\angstrom}$ (Cu K$\mathrm{\alpha}$ radiation), $\Delta(2\theta)$ FWHM of the reflex at position $2\theta$, the lower limit of the vertical expanse of the crystallites, $l\mtfg{z}$, can be estimated. On the one hand, for room-temperature PLD-grown NiO, $l\mtfg{z}$ amounts to approximately \SI{30}{\nano\meter}. On the other hand, the absence of the reflex for the sputtered films implies grains smaller than about \SI{5}{\nano\meter}. In order to discriminate between effects arising from the crystalline structure and from the true bulk transport properties of NiO, PLD growth at an elevated temperature of \SI{300}{\degreeCelsius} was studied as well. Such films grow in a highly (111) oriented fashion, with a grain size, according to Eqn.~(\ref{eq:scherrer}), of around \SI{150}{\nano\meter} being of the same order as the film thickness of this sample.
	
	\subsection{Doping concentrations}

	Intrinsic doping of NiO is achieved by suppling excess oxygen during film growth. This leads to a high density of nickel vacancies (V\tfg{Ni}) which is the dominant intrinsic point defect under oxygen-rich ambient conditions \cite{Lany.2007}. In the experiments reported here, this is achieved using both deposition methods by choosing a high oxygen background pressure ($p\mtfg{O2}=\SI{0.1}{\milli\bar}$ and \SI{0.018}{\milli\bar} for PLD and sputtering, respectively). In order to determine the resulting carrier density for the particular growth conditions, films were grown on top of highly fluorine-doped tin oxide (FTO) layers \footnote{FTO-coated glass substrates were supplied by Calyxo GmbH, Germany.} ($\sigma\mtfg{FTO} = \SI{2.5e3}{\siemens\per\centi\meter}$) which induces a significant depletion region in the NiO layers. On these structures, capacitance-voltage ($CV$) measurements were conducted to determine the net doping density $N\mtfg{net}$ of the films; Hall effect measurements can usually not be employed to NiO due to the absence of free carriers. The values for $N\mtfg{net}$ are given in Table~\ref{tab:samples}. In the further course of this report, $N\mtfg{net}$ is interpreted as being equal to the carrier density. Tab.~\ref{tab:samples} demonstrates that reactive DC magnetron sputtering produces films with the highest net doping ($N\mtfg{net}\approx$\SI{e19}{\per\cubic\centi\meter}), whereas high-temperature PLD-grown films have the lowest doping density (\SI{e18}{\per\cubic\centi\meter}).
	
	\begin{table}
		\caption{Parameters of NiO films fabricated by different methods, namely reactive DC magnetron sputtering, ( "ms-NiO") and PLD at room temperature and at \SI{300}{\degreeCelsius} (RT-PLD- and HT-PLD-NiO, respectively): grain size $l\mtfg{z}$, net doping density $N\mtfg{net}$ from $CV$ measurements, and DC conductivity $\sigma\mtfg{dc}$ at room temperature.}
		\label{tab:samples}
		\begin{tabular}{lccc} \toprule
			sample & $l\mtfg{z}$ (\si{\nano\meter}) & $N\mtfg{net}$ (\si{\per\cubic\centi\meter}) & $\sigma\mtfg{dc}$ (\si{\siemens\per\centi\meter}) \\ \colrule
			ms-NiO & $< 5$ & \num{1e19} & \num{9e-4} \\
			RT-PLD-NiO & $\approx \num{30}$ & \num{4e18} & \num{3e-6} \\
			HT-PLD-NiO & $\ge \num{150}$ & \num{1e18} & \num{3e-8} \\ \botrule
		\end{tabular}
	\end{table}
	
	\subsection{Current-voltage characterization}
	
	\begin{figure}
		\centering
		\includegraphics[width=\columnwidth]{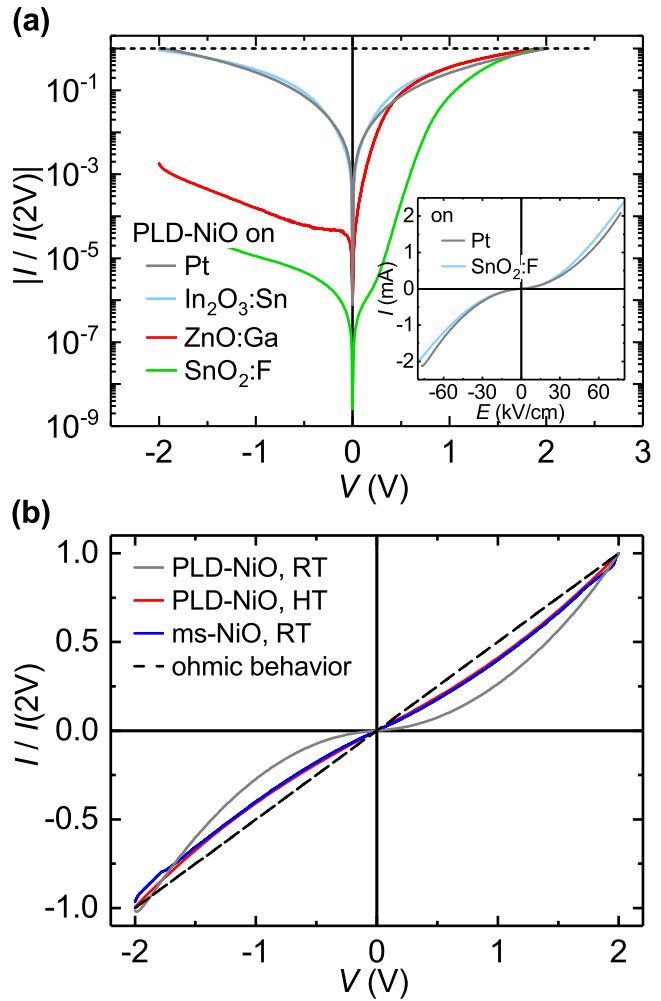}
		\caption{(a) Typical current-voltage characteristics of room-temperature PLD-grown NiO on top of different conductive materials; absolute currents in log scale, normalized to currents at $V=\SI{2}{\volt}$, for all back contacts. Inset: currents for NiO on Pt and FTO as function of the applied DC electric field due to different NiO thicknesses (on ITO: \SI{100}{\nano\meter}, on Pt: \SI{260}{\nano\meter}). (b) normalized currents for differently grown NiO layers, and ohmic characteristic for comparison.}
		\label{fig:IV_different_back_contacts}
	\end{figure}

	The starting point for the investigations in this section is the general observation of non-ohmic DC transport in NiO thin films, even in the absence of a depletion layer. Fig.~\ref{fig:IV_different_back_contacts}(a) shows typical current-voltage relationships for NiO layers sandwiched between different conductive materials. Specifically, fluorine-doped tin oxide (FTO), tin-doped indium oxide (ITO), gallium-doped zinc oxide (GZO) (the latter two grown in-house by PLD), and Pt were used as bottom contacts, respectively. All samples were capped with a Pt top electrode. It can be seen that FTO and GZO back electrodes induce a depletion layer located almost exclusively within the NiO (hole density ${p = N\mtfg{net} = \SI{4e18}{\per\cubic\centi\meter}}$), due to the high electron densities ($n>\SI{5e19}{\per\cubic\centi\meter}$, determined by Hall effect measurements) in GZO and FTO layers. This depletion layer causes a significant current rectification. In the case of ITO contacts, the rectification is hardly visible, and for the Pt back electrode the current-voltage characteristics are almost symmetric, indicating the absence of a depletion region. However, a clear deviation from a linear current-voltage relationship is present also in these cases, as shown in the inset of Fig.~\ref{fig:IV_different_back_contacts}(a). It can be seen that the $IV$ characteristics (apart from a weak asymmetry in the case of ITO due to slight depletion) show very similar behavior, with a conductivity that increases with applied electric field. The similarity between the two different contact materials and the independence of the NiO thickness implies that the observed behavior is intrinsic to the NiO layers and not an effect of a voltage-dependent carrier injection at the contacts. 
	
	Fig.~\ref{fig:IV_different_back_contacts}b displays normalized current-voltage characteristics for NiO layers grown by the different methods used in this work. The deviation from ohmic behavior (indicated by the dashed black line) can be seen for all samples, but is most pronounced for room-temperature PLD-grown layers. For this reason, these samples were chosen for further investigations of transport properties. It should be noted that, although the characteristics match well qualitatively, the current densities are different for those three samples due to widely varying conductivities. Values for $\sigma\mtfg{dc}$, obtained by numerical differentiation of the current-voltage characteristics at a DC voltage of \SI{0}{\volt}, are provided in Tab.~\ref{tab:samples}. It can be stated that there seems to be a strong nonlinear dependence of $\sigma\mtfg{dc}$ on $N\mtfg{net}$, indicating that the role of the V\tfg{Ni} acceptors probably goes beyond contributing extra charge carriers to the electronic system.
	
	Non-linear current transport in NiO has been reported before, e.g. on Pt/NiO/Pt \cite{Seo.2005} and ITO/NiO structures \cite{Chang.2006}. This behavior has been interpreted by Chang \textit{et al.} in the framework of the space charge-limited current (SCLC) model \cite{Rose.1955} which explicitly takes into account long transient times for charge carriers travelling through a low-mobility insulator after injection by ohmic contacts. These authors found an approximate $I\propto V^2$ relationship, indicating that the transport within the NiO layer is limited by hole trapping into electronic states with a single activation energy.
	
\begin{figure}
		\centering
		\includegraphics[width=\columnwidth]{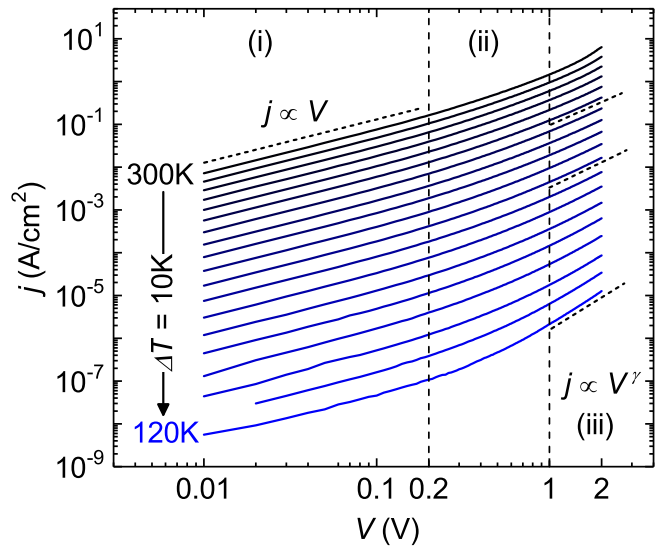}
		\caption{Current density-voltage characteristics for one contact in double-logarithmic scale for temperatures between \SI{300}{\kelvin} and \SI{120}{\kelvin}.}
		\label{fig:IUT_loglog}
	\end{figure}

	In Fig.~\ref{fig:IUT_loglog}, the temperature dependence of the $jV$ characteristics of Pt/(RT)PLD-NiO/Pt structures is shown. Three different regions of the curves can be distinguished on a double-logarithmic scale: (i) ohmic conduction below \SI{0.2}{\volt} $j \propto V$, (ii) a transition region where the slope of the curves increases, and (iii) a power-law region $j \propto V\mhog{\gamma}$ with a temperature-dependent voltage exponent $\gamma$ taking on values close to 2. It should be noted that in the data presented here, a slight exponential contribution to the current densities for temperatures above \SI{260}{\kelvin} is detected in the voltage regime (iii) owing to a weak hole barrier located beneath the metallic Pt capping layer. This behavior is observed when the capping layer is not deposited on top of the NiO immediately after semiconductor growth. We therefore think that it can be attributed to an out-diffusion of the mobile Ni vacancies in the vicinity of the NiO surface, thereby producing a downward bending of the bands that is conserved also after Pt deposition. We have tested this hypothesis by using different combinations of the noble metals Au, Pd and Pt for both back and front contacts and found that the occurrence of a slight rectification is only dependent on the time between NiO and capping layer deposition, with the depletion layer always located below the top electrode. For temperatures  $ \le \SI{260}{\kelvin}$, however, the current densities are symmetrical with respect to the applied voltage, and therefore are probably determined solely by transport in the neutral NiO bulk material.
	
\begin{figure}
		\centering
		\includegraphics[width=\columnwidth]{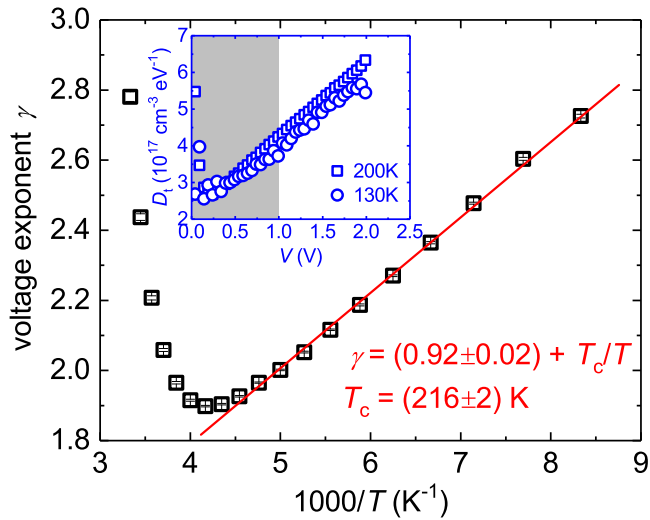}
		\caption{Temperature dependence of the voltage exponent $\gamma$, and linear fit according to Eq.~(\ref{eq:voltageExponentTemp}). Inset: trap distribution calculated for $T = \SI{200}{\kelvin}$ and \SI{130}{\kelvin} according to Eq.~(\ref{eq:trapDistrFromjV}); gray area: ohmic and transition region where unphysical results are likely obtained.}
		\label{fig:IUT_voltage-exponent}
	\end{figure}

	The temperature dependence of the voltage exponent $\gamma$ is shown in Fig.~\ref{fig:IUT_voltage-exponent}. It is evident that below \SI{220}{\kelvin} $\gamma$ is linearly dependent on $T\mhog{-1}$, increasing from a value of \num{1.92} at \SI{220}{\kelvin} to \num{2.72} at \SI{120}{\kelvin}. This behavior can be consistently explained in the framework of SCLC, taking into account trap states distributed over a certain energy range as proposed by Rose \cite{Rose.1955}. It can be shown that in this case, the relation
	
	\begin{equation}
	j \propto V\mhog{1 + \frac{T\mtfg{c}}{T}}
	\label{eq:voltageExponentTemp}
	\end{equation}
	
	holds, with $T\mtfg{c}$ a critical temperature characteristic for the broadness of the trap distribution. Fitting the data $\gamma(T\mhog{-1})$ with a linear model yields ${\gamma = \num{0.92} + \frac{\SI{216}{\kelvin}}{T}}$, which is in agreement with Eq.~(\ref{eq:voltageExponentTemp}). It can be concluded that the transport in NiO can be characterized as being due to holes trapped in a distribution of states. Rose \cite{Rose.1955} provided a method which allows to determine the trap distribution on the basis of $jV$ characteristics under the condition that $\gamma \ge 2$:
	
	\begin{equation}
	D\mtfg{t} =  \frac{1}{dA} \frac{CV}{e k\mtfg{B}T} \left(\frac{V}{j} \frac{\diff j}{\diff V}-1 \right)\mhog{-1}
	\label{eq:trapDistrFromjV}
	\end{equation}
	
	with $d$ and $A$ being the thickness and area of the contact, respectively, and $C$ its capacitance (\SI{85}{\pico\farad}, as determined from quasi-static capacitance-voltage measurements). Eq.~(\ref{eq:trapDistrFromjV}) gives the number density of traps per unit energy interval. This calculation has been performed for two different temperatures, namely \SI{200}{\kelvin} and \SI{130}{\kelvin}; the results are shown in the inset of Fig.~\ref{fig:IUT_voltage-exponent}. It can be seen that the density of traps determining the admitted current density increases linearly with applied voltage. It is also worth noticing that the value of the probed density of states does not seem to depend on temperature in this range. The observed decrease of the electric conductivity can therefore be entirely ascribed to a drop of the charge carrier mobility. 
	
	Because the hole in NiO occupies a Zhang-Rice bound state, implications for the electric current transported by these states should be considered \cite{Taguchi.2008}. Due to the similarity of the current-voltage characteristics of differently deposited NiO layers in our labs, and also the ones reported in the literature, we are convinced that the non-ohmic behavior is specific to the intrinsic transport of ZR states. The following scenario is suggested as a possible explanation: The formation of the ZR states depends the presence of spins localized on Ni $3d$ states on the one hand, and of holes (e.g, in the O $2p$ band, around V\tfg{Ni} sites etc.) on the other hand. Because the application of an external voltage injects additional holes, their density is enhanced, thereby leading to a more pronounced ZR band. Since the amount of injected carriers scales linearly with the applied voltage, a linear correlation between $V$ and the ZR density of states can be expected, which is exactly what is observed (inset in Fig.~\ref{fig:IUT_voltage-exponent}). This process might be of importance for the resistive switching behavior of NiO thin films, in which the conductivity changes abruptly upon applying a critical set voltage. This has already been attributed to the injection-induced formation of conductive filaments  \cite{Son.2008,Ielmini.2011}. It should be noted that this effect was never observed on NiO films investigated in this work because of their comparatively high conductivities. Resistive switching is usually observed in films that exhibit insulating behavior in their pristine state.
	
	\begin{figure}
		\centering
		\includegraphics[width=\columnwidth]{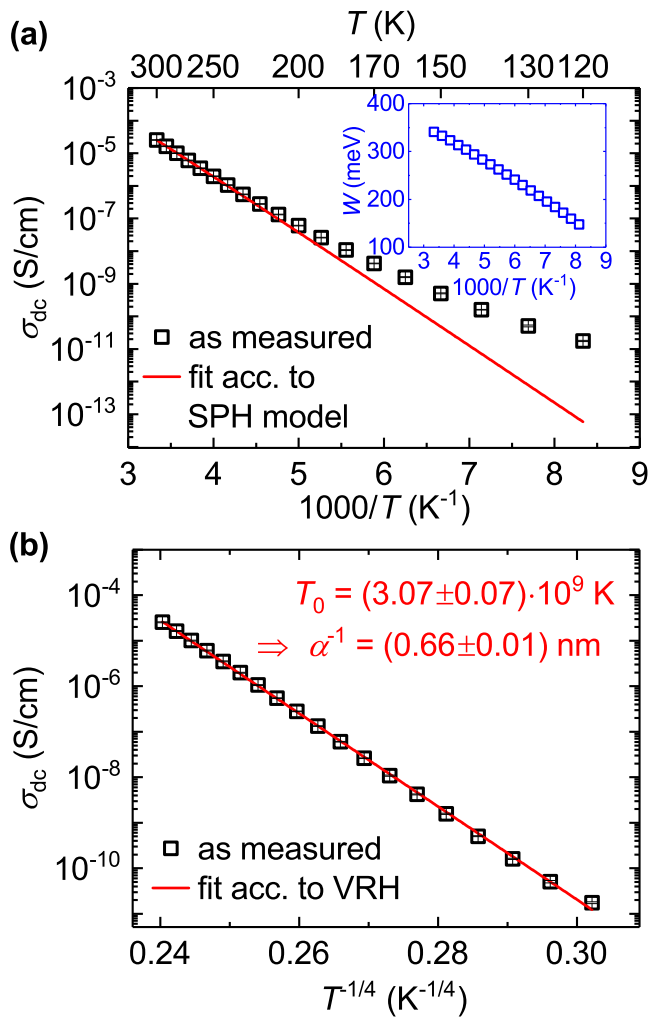}
		\caption{Temperature dependence of the DC conductivity of PLD-grown NiO, obtained by numerical differentiation of the data from Fig.~\ref{fig:IUT_loglog}, and fits according to (a) the small polaron hopping model and (b) Mott variable range hopping. Inset in (a) shows the temperature dependence of the apparent hopping energy.}
		\label{fig:sigmaDC_SPH+VRH}
	\end{figure}

	Fig.~\ref{fig:sigmaDC_SPH+VRH}a shows the temperature dependence of the conductivity as extracted from the data shown in Fig.~\ref{fig:IUT_loglog} by numerical differentiation, for $V\mtfg{dc} = \SI{0}{\volt}$. As can be seen, the conductivity decreases when the temperature is lowered. Furthermore, the thermal activation deviates from a single Arrhenius type as is evident from a kink of the curve at around \SI{200}{\kelvin}. To describe the $T$-dependence of the conductivity $\sigma\mtfg{dc}$ of NiO, a transport model based on the hopping of "free" small polarons has been used in recently published works \cite{Zhang.2018,Liu.2019}. Within this model, the simplified expression 
	
	\begin{align}
	\sigma\mtfg{dc} &= \sigma_0\exp\left(-\frac{W\mtfg{H}}{k\mtfg{B}T}\right) \nonumber \\ 
    \text{with}	\qquad \sigma_0 &= \frac{Nea\mhtg{0}{2}\nu\mtfg{ph}}{k\mtfg{B}T} 
	\label{eq:SPH_conductivity}
	\end{align}
	
	holds \cite{Austin.1969,Bosman.1970} with the polaron hopping activation energy $W\mtfg{H}$, carrier density $N$, $a_0$ lattice constant and $\nu\mtfg{ph}$ optical phonon frequency. In Fig.~\ref{fig:sigmaDC_SPH+VRH}a, fits according to Eqn.~\ref{eq:SPH_conductivity} are shown. It is obvious that this model can reproduce the data only for temperatures above approximately \SI{200}{\kelvin}, although it must be said that in this region, also a single Arrhenius-type activation would be sufficient. For lower temperatures, the SPH model predicts considerably smaller values of $\sigma\mtfg{dc}$ than experimentally determined, owing to a decrease of the activation energy with falling temperature (shown in the inset of Fig.~\ref{fig:sigmaDC_SPH+VRH}a). Such a drop in $W\mtfg{H}$ is included in more elaborate models of small polaron hopping \cite{Bottger}. Therein, the drop evolves at temperatures of about one half of the Debye temperature $\frac{\theta\mtfg{D}}{2}$, which is approximately \SI{200}{\kelvin} for NiO \cite{Allen.1954}. Apart from that, however, a renewed \textit{increase} of the conductivity $\sigma\mtfg{dc}$ is to be expected at further decreasing temperatures due to the onset of polaronic band conduction. This has not been observed in the NiO samples investigated in this work down to temperatures of \SI{20}{\kelvin}. Moreover, the extracted carrier mobilities are extremely low -- at room temperature, $\mu = \frac{\sigma\mtfg{dc}}{eN\mtfg{net}} \approx \SI{e-6}{\square\centi\meter\per\volt\per\second}$, assuming the carrier density from $CV$ measurements, $N\mtfg{net} = \SI{4e18}{\per\cubic\centi\meter}$. We suggest to discard hopping conduction of "free" small polarons as current transport mechanism in doped NiO. The fact that the carriers are of polaronic character, however, is not questioned here, because this is an inherent feature of the bound ZR state \cite{Bala.1994}. We propose to interpret the data in the framework of the polaronic impurity hopping conduction as described by Schnakenberg \cite{Schnakenberg.1968}. Within this model, the carriers are small polarons transported by hopping, but they cannot move to any lattice site. Instead, their propagation is limited to defect sites to which they are tightly bound -- a situation readily applicable to carriers occupying ZR bound states. In this model, the conductivity follows the same temperature dependence as in the "free" small polaron hopping, exhibiting a significant drop of activation energy at about $\frac{\theta\mtfg{D}}{2}$. The model differs only at low temperatures where $\sigma\mtfg{dc}$ continues to decrease according to $\exp\left(-\frac{A}{k\mtfg{B}T}\right)$, where $A$ is an activation energy of the order of the mean energy separation $\Delta\epsilon$ between neighboring defect sites. Band conduction is suppressed in this case due to the strong localization and large inter-site separation. This model has already been applied to Li-doped NiO \cite{Springthorpe.1965,Mott.1968}, and it is applicable to intrinsically (V\tfg{Ni}-) doped NiO. In this case, however, the term "polaronic inter-acceptor hopping" is more appropriate because V\tfg{Ni} defects are not impurities in a chemical sense.
	
	Carrier hopping between defect states randomly distributed in space and energy (in a sufficiently large range $\Delta\epsilon$) can also be described within the model of variable range hopping (VRH) developed by Mott \cite{Mott.1968a}. There, the relation
	
	\begin{equation}
	\sigma\mtfg{dc} = \sigma_0 \exp\left(-\left[\frac{T_0}{T}\right]\mhog{\frac{1}{4}}\right)
	\label{eq:sigmaDC_VRH}
	\end{equation}
	
	holds, with a characteristic temperature 
	
	\begin{equation}
	T_0 \approx \frac{20\alpha^3}{k\mtfg{B}N(\epsilon\mtfg{F})}
	\end{equation}
	
	which is determined by the localization length $\alpha^{-1}$ of the carriers and the density of states at the Fermi energy $N(\epsilon\mtfg{F})$. In Fig.~\ref{fig:sigmaDC_SPH+VRH}b, the measured DC conductivity is plotted logarithmically against $T\mhog{-\frac{1}{4}}$ together with a fitting curve according to Eqn.~(\ref{eq:sigmaDC_VRH}), and it can be seen that the VRH model reproduces the data over the entire temperature range investigated here. The characteristic temperature amounts to \SI{3.07\pm0.07 e9}{\kelvin}. If one takes the acceptor density of states $N(\epsilon\mtfg{F}) = D\mtfg{t} =  \SI{2.6e17}{\per\cubic\centi\meter\per\electronvolt}$, as obtained from the $jV$ characteristics by extrapolating the density of states in Fig.~\ref{fig:IUT_voltage-exponent} to $V=\SI{0}{\volt}$, the localization length of a polaron in a ZR state amounts to $\alpha\mhog{-1} = \SI{0.66\pm0.01}{\nano\meter} \approx \num{1.6} a_0$.\\
	
	One might object to using the VRH model in the entire temperature range because this model has originally been developed for transport at low temperatures only. However, B\"{o}ttger and Bryksin \cite{Bottger} formulate a criterion that can be employed for justifying the use of this model also at higher temperatures. When the energetic distribution of the acceptor states $\Delta\epsilon$ is narrow, such that
	
	\begin{equation}
	\alpha N\mtfg{net}\mhog{-\frac{1}{3}} \gg \frac{\Delta\epsilon}{k\mtfg{B}T},
	\label{eq:crit_NNH}
	\end{equation}
	
	holds true, next-neighbor hopping occurs. When Eqn.~(\ref{eq:crit_NNH}) is not fulfilled, VRH is favored. In the present case, ${N\mtfg{net} = \SI{4e18}{\per\cubic\centi\meter}}$, $\alpha = \SI{1.5e9}{\per\meter}$, and $\Delta\epsilon \approx \SI{200}{\milli\electronvolt}$ (as extracted from the apparent activation energy of $\sigma\mtfg{dc}$ at low temperatures), such that at room temperature ${\alpha N\mtfg{net} = \num{9.4} \not\gg \frac{\Delta\epsilon}{k\mtfg{B}T} \approx 8}$. Thus, the VRH model can be applied at least up to room temperature. It can be comprehended as a limiting case of the inter-acceptor hopping model for large energy spread $\Delta\epsilon$ of the defect states. Such a broad energetic distribution can be expected due to the high degree of disorder in the RT-deposited NiO films leading to the formation of other defects or defect complexes, some of which are donor-like. This causes a large degree of compensation. In fact, for the occurrence of charge transport, partial compensation is always necessary, because otherwise all acceptors were charge-neutral, and no free target states were available for hopping.
	
\subsection{BDS measurements}

\begin{figure*}
	\centering
	\includegraphics[width=1.8\columnwidth]{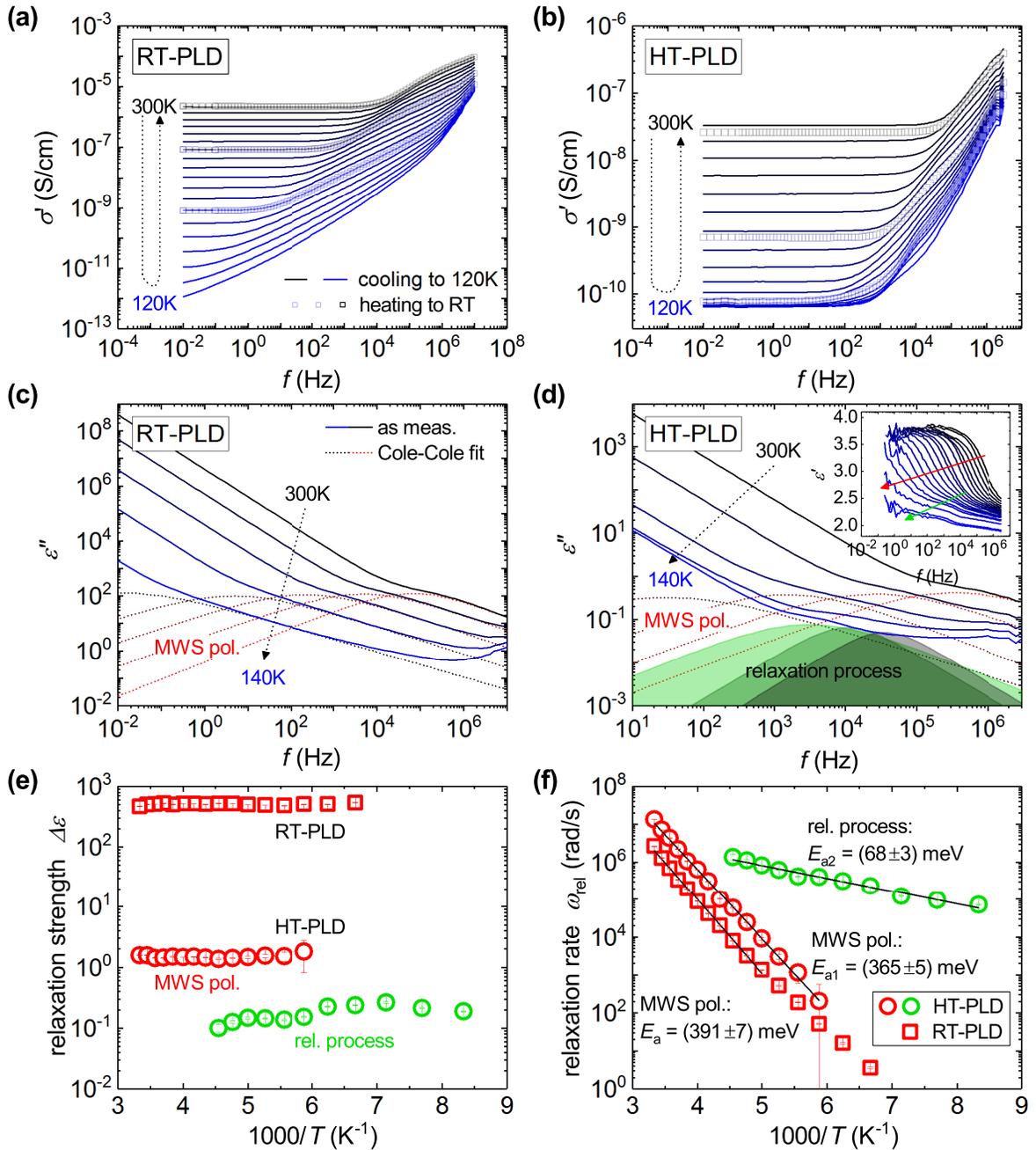}
	\caption{Frequency and temperature dependence of (a,b) the real part $\sigma'$ of the complex conductivity $\sigma\mhog{*} = \sigma' + \imath \sigma''$, and (c,d) the imaginary part $\varepsilon''$ of the complex dielectric function $\varepsilon\mhog{*} = \varepsilon' - \imath \varepsilon''$, for (a,c) the room temperature- and (b,d) high temperature-deposited NiO film for selected temperatures, together with fits of the relaxation processes according to Eqn.~(\ref{eq:colecole}). Inset in (d): real part of dielectric function for all temperatures, arrows indicating steps due to the two relaxation processes. (e) relaxation strengths $\Delta \varepsilon$ and (f) relaxation rates $\omega\mtfg{rel} = \frac{2\pi}{\tau\mtfg{rel}}$ of the identified processes.}
	\label{fig:BDS_data_RT+HT_cooling}
\end{figure*}

In this section, the temperature-dependent behavior of the complex conductivity $\sigma\mhog{*} = \sigma' + \imath \sigma''$ of NiO thin films is discussed, comparing the sample grown by PLD at room temperature ("RT-PLD") and the one at \SI{300}{\degreeCelsius} ("HT-PLD"). Fig.~\ref{fig:BDS_data_RT+HT_cooling}a and b show the real part of the conductivity, $\sigma'$, measured by broad-band dielectric spectroscopy (BDS) between room temperature and \SI{120}{\kelvin} and in a frequency range from \SI{e-2}{\hertz} to \SI{e7}{\hertz} (RT-PLD) or \SI{3e6}{\hertz} (HT-PLD). In the latter case, the data obtained above \SI{3e6}{\hertz} were discarded due to the occurrence of inductance-induced artifacts. Measurements were taken while cooling to \SI{120}{\kelvin} as well as during heating back to room temperature in order to check for the reproducibility. Measurements were done at a DC voltage of \SI{0}{\volt}, the amplitude of the AC voltage was set to \SI{10}{\milli\volt} such that the response of the sample is measured in the ohmic voltage region. 

Two different features dominate the behavior of $\sigma'$: (i) a frequency-independent contribution at low frequencies appearing as a plateau and interpreted as the DC conductivity limit $\sigma\mtfg{dc}$, as well as (ii) a dispersive part where $\sigma'$ increases with frequency. In the case of the RT-deposited film, $\sigma\mtfg{dc}$ gradually decreases over more than 6 orders of magnitude within the investigated temperature range, whereas for the HT-NiO it drops by approximately 3 orders of magnitude down to approximately \SI{200}{\kelvin} and remains almost constant below that temperature.

The complex dielectric function $\varepsilon\mhog{*}$ can be calculated from the complex conductivity $\sigma\mhog{*}$ via the relation

\begin{equation}
\varepsilon\mhog{*} =  \frac{1}{\imath \omega \varepsilon_0}\sigma\mhog{*}.
\end{equation}

In Fig.~\ref{fig:BDS_data_RT+HT_cooling}c and d, the imaginary part of $\varepsilon\mhog{*}$, ${\varepsilon'' = \frac{1}{\omega \varepsilon_0}\sigma'}$ is plotted. It becomes evident that the dispersion of $\sigma'$ is due to the occurrence of dielectric relaxation phenomena, expressed as peaks or shoulders on the high-frequency side of the DC conductivity term ($\propto \omega\mhog{-1}$). These relaxation processes can be described with the Cole-Cole equation \cite{Kremera}

\begin{align}
\varepsilon\mhtg{CC}{*}(\omega) &= \varepsilon'\mtfg{CC}(\omega) + \imath \varepsilon''\mtfg{CC}(\omega) \nonumber \\
& = \varepsilon\mtfg{\infty} + \frac{\varepsilon\mtfg{s}-\varepsilon\mtfg{\infty}}{1+\left(\imath \omega \tau\mtfg{rel}\right)\mhog{1-b}},\nonumber \\
\varepsilon'\mtfg{CC} &= \varepsilon\mtfg{\infty} + \frac{\varepsilon\mtfg{s}-\varepsilon\mtfg{\infty}}{2} \nonumber \\
&\times \left[1-\frac{\sinh\left((1-b)\ln(\omega\tau\mtfg{rel})\right)}{\cosh\left((1-b)\ln(\omega\tau\mtfg{rel})\right)+\cos b \pi /2}\right], \nonumber \\ 
\varepsilon''\mtfg{CC} &= \frac{\varepsilon\mtfg{s}-\varepsilon\mtfg{\infty}}{2} \frac{\cos b \pi /2}{\cosh \left((1-b)\ln(\omega\tau\mtfg{rel})\right)+\sin b \pi /2}
\label{eq:colecole}
\end{align}

which is a Debye-type function with symmetric broadening. The parameters $\Delta\varepsilon = \varepsilon\mtfg{s} - \varepsilon\mtfg{\infty}$, $b \in [0,1)$ and $\tau\mtfg{rel}$ represent the relaxation strength, relaxation process broadening and relaxation time, respectively. Based on Eqn.~(\ref{eq:colecole}), and by adding the DC conductivity term ($\varepsilon'' = \varepsilon''\mtfg{CC} + \frac{\sigma\mtfg{dc}}{\omega \varepsilon_0}$), the $\varepsilon''$ data have been fitted to extract the respective relaxation process parameters and the DC conductivity. These fits are included in Figs.~\ref{fig:BDS_data_RT+HT_cooling}c and d as color-shaded areas. The fit parameters $\Delta\varepsilon$ and $\omega\mtfg{rel} = \frac{2\pi}{\tau\mtfg{rel}}$ and their temperature dependence are shown in Figs.~\ref{fig:BDS_data_RT+HT_cooling}e and f. The data for the broadening parameter $b$ is not presented.

In the case of the RT-grown film with high structural disorder, only one relaxation process is visible. It possesses a generally high relaxation strength $\Delta\varepsilon\approx 500$ which is almost independent on temperature (Fig.~\ref{fig:BDS_data_RT+HT_cooling}e). The film with low disorder exhibits two relaxation processes, the stronger of which appears between room temperature and about \SI{170}{\kelvin}, while the weaker process is only visible below \SI{230}{\kelvin}. For reasons explained in the following, the relaxation process in the RT-deposited sample and the stronger one in the HT-deposited film are probably due to the same mechanism; this process will be analyzed first.

For a process caused by relaxation of molecular dipoles of density $N\mtfg{dip}$ and dipole moment $\mu\mtfg{dip}$, $\Delta\varepsilon$ at a certain temperature can be estimated by the relation

\begin{equation}
\Delta\varepsilon \approx \frac{1}{\varepsilon_0}\frac{\mu\mtfg{dip}^2}{k\mtfg{B}T}N\mtfg{dip}.
\label{eq:dipole_relaxation}
\end{equation}

If one assumes that the dipoles consist of two charges $\pm e$ at a distance of one NiO lattice spacing $a_0$, and the dipole density is given by the trapped-hole density, $N\mtfg{dip} \approx \SI{1e18}{\per\cubic\centi\meter}$, the corresponding relaxation strength at room temperature, according to Eqn.~(\ref{eq:dipole_relaxation}) would be of the order of \num{0.1} which is three orders of magnitude lower than the $\Delta\varepsilon$ observed in the RT-deposited film and a factor of 10 lower than in the HT-NiO sample. In addition, Eqn.~(\ref{eq:dipole_relaxation}) implies a decrease of $\Delta\varepsilon$ with increasing temperature; however, such a decrease is absent in the measurements. Therefore, a molecular relaxation mechanism can be excluded as assignment for this relaxation process. Instead, an explanation is provided on the basis of the crystalline structure of the investigated films, which causes a spatial inhomogeneity of the electrical conductivity. Such mesoscopic inhomogeneities are known to cause interfacial polarization effects, subsumed under the term \textit{Maxwell-Wagner-Sillars (MWS) polarization} \cite{Kremer}. Models for this type of effect typically consider particles with a high electrical conductivity embedded in a matrix of lower conductivity, which cane be comprehended as conductive NiO grains separated by insulating grain boundaries here. Consequently, an AC electric field induces a spatial separation of charge carriers within the grains. A characteristic of an MWS process is an abnormally high and temperature-independent relaxation strength ($\Delta \varepsilon \gg 1$, determined by the conductivity contrast between the two components and the filling ratio of the particles).

Typical for disordered materials with a distinct DC conductivity plateau is a dielectric relaxation process visible at frequencies which are to high for the mechanism causing the DC conductivity (usually a molecular hopping process) \cite{Dyre.2000}. The correlation between $\sigma\mtfg{dc}$ and the relaxation strength and rate is described by the so-called \textit{Barton-Nakajima-Namikawa (BNN) relation}:

\begin{equation}
\sigma\mtfg{dc} = p\cdot \varepsilon_0 \Delta\varepsilon\, \omega\mtfg{rel}
\label{eq:BNN}
\end{equation}

\begin{figure}
	\centering
	\includegraphics[width=\columnwidth]{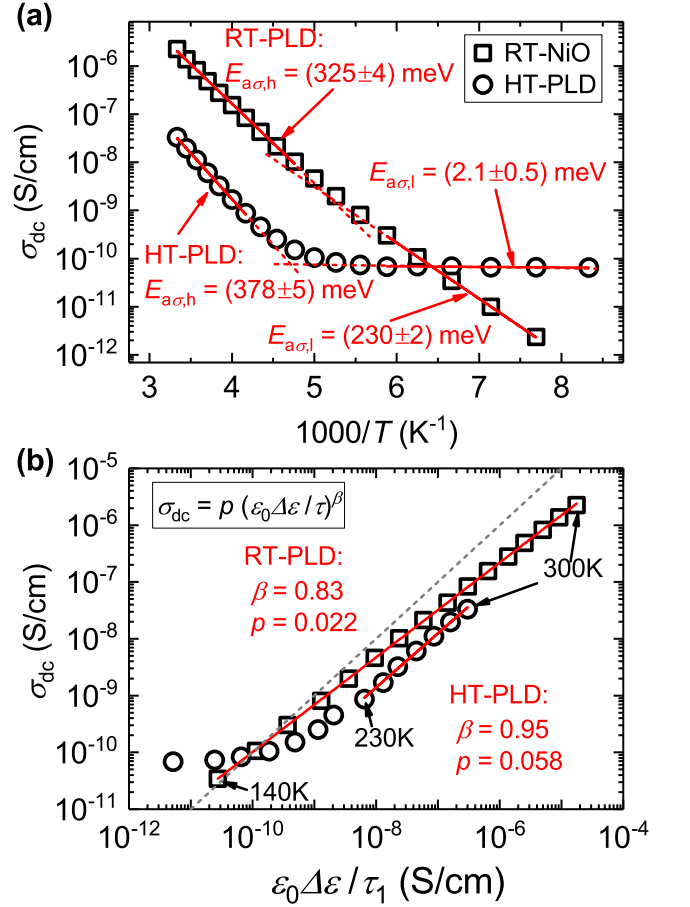}
	\caption{(a) Temperature dependence of the DC conductivity for the RT and HT-deposited sample as extracted from the frequency-dependent measurements, with Arrhenius fits and respective activation energies. (b) Barton-Nakajima-Namikawa (BNN) relation for the strong relaxation process observed in both samples; gray line indicates the ideal (${p = 1, \beta = 1}$) relationship.}
	\label{fig:sigmaDC+BNN}
\end{figure}

Here, $p$ is a numerical constant of the order of unity, typically between \num{0.5} and 10, and independent of temperature. Because the $T$-dependence of $\Delta\varepsilon$ is typically much weaker than that of $\sigma\mtfg{dc}$, Eqn.~(\ref{eq:BNN}) implies that $\sigma\mtfg{dc} \propto \omega\mtfg{rel}$. In addition, if the DC conductivity is temperature-activated with a particular activation energy, $\omega\mtfg{rel}$ should show the same dependence. This has been checked for the investigated NiO films and shown in Fig.~\ref{fig:sigmaDC+BNN}a and b. In the temperature region between \SI{300}{\kelvin} and about \SI{240}{\kelvin}, where $\sigma\mtfg{dc}$ shows a temperature-activated behavior, the activation energies are \SI{325}{\milli\electronvolt} and \SI{378}{\electronvolt} for the RT and HT sample, respectively, which is close to the activation energies for the relaxation process in question (\SI{391}{\milli\electronvolt} and \SI{365}{\milli\electronvolt}, see Fig.~\ref{fig:BDS_data_RT+HT_cooling}f). In Fig.~\ref{fig:sigmaDC+BNN}b, it can be seen that the BNN relation between the DC conductivity and the relaxation parameters is fairly satisfied in the range where $\sigma\mtfg{dc}$ is temperature-activated -- for the HT-NiO sample, that excludes the low-temperature regime. The relationship between $\sigma\mtfg{dc}$ and the "reduced relaxation strength" $\varepsilon_0\Delta\varepsilon\,\omega\mtfg{rel}$ is, however, slightly sublinear, and the parameter $p$ takes on values of the order of $\num{e-2} \ll 1$. Such a low value of the BNN parameter $p$ is far outside the range expected for a microscopic origin of the relaxation process. The similarity of the activation energies of $\sigma\mtfg{dc}$ and $\omega\mtfg{rel}$ is, on the other hand, a typical signature of an MWS polarization process \cite{Kremer}. Within a simplified model that considers spherical filler particles (volume fraction $\varphi\mtfg{f}$) of conductivity $\sigma\mtfg{f}'$ inside a matrix of conductivity $\sigma\mtfg{m}'$ with $\sigma\mtfg{m}' \ll \sigma\mtfg{f}'$, filling fraction $\varphi\mtfg{m} = 1 - \varphi\mtfg{f}$ and an identical real part of the dielectric function $\varepsilon'$ for both phases, it can be shown that a Debye-type relaxation occurs \cite{Kremer}, with 

\begin{align}
\Delta\varepsilon &\approx 3 \varepsilon' \left(\frac{\sigma\mtfg{f}'}{\sigma\mtfg{m}'}\right)^2 \frac{\varphi\mtfg{f}\left(1-\varphi\mtfg{f}\right)}{2+\varphi\mtfg{f}}, \nonumber \\ 
\tau\mtfg{rel} &\approx 3 \varepsilon_0 \varepsilon' \frac{1}{\sigma\mtfg{f}'\left(1-\varphi\mtfg{f}\right)}.
\label{eq:MWS_parameters}
\end{align}

These equations demonstrate that, on the one hand, the relaxation strength is determined by the squared ratio of the conductivities $\sigma\mtfg{f}$ and $\sigma\mtfg{m}$. The temperature dependencies of both are similar and therefore roughly cancel out, and $\Delta\varepsilon$ becomes only weakly temperature-dependent. On the other hand, the relaxation rate $\omega\mtfg{rel} \propto \tau\mtfg{rel}\mhog{-1}$ is directly proportional to $\sigma\mtfg{f}$, and therefore a BNN relation-like correlation is obtained. Eqns.~\ref{eq:MWS_parameters} are, however, only valid under rather strict conditions regarding the filling fraction $\varphi\mtfg{f}$ and the shape of the filler particles, and are therefore only used for constructing the qualitative arguments and not for quantitative analysis here. In any case, it can be stated that the difference in relaxation strength of the intense process is caused by a lower degree of disorder in the HT-grown film as compared to the RT-grown one, causing a reduced spatial inhomogeneity of the conductivity.

One may ask whether this polarization process renders the previously mentioned $CV$ measurements unreliable, because it causes a large share of the RT-grown sample's capacitance at and below audio frequencies around room temperature, where these measurements were recorded. The capacitance of the depleted films used for $CV$ characterization is in the range of \SI{500}{\pico\farad}, which is a factor of 6 larger than the value for the undepleted films (about \SI{85}{\pico\farad}). The overestimation of the net doping density caused by the MWS process therefore amounts to \SI{34}{\percent} of its value, because $N\mtfg{net} \propto C^2$. The net doping of the RT-grown film is therefore approximately \SI{2.6e18}{\per\cubic\centi\meter}.	

Next, the weaker relaxation process observed in the low-disorder sample is considered, which is only visible below \SI{230}{\kelvin}. Above this temperature, it is potentially masked by the more intense MWS process, although there seems to be an abrupt qualitative change (which can be well seen in $\varepsilon'$, see inset of Fig.~\ref{fig:BDS_data_RT+HT_cooling}d) exactly at \SI{240}{\kelvin} (Fig.~\ref{fig:BDS_data_RT+HT_cooling}d). This process has a much lower relaxation strength $\Delta\varepsilon \approx \num{0.1}$ than the MWS process (see Fig.~\ref{fig:BDS_data_RT+HT_cooling}e) which also shows a slight decrease with increasing temperature. Both facts are in accordance with a microscopic hopping process. The relaxation rate of the process is activated with an energy of \SI{68\pm3}{\milli\electronvolt} (Fig.~\ref{fig:BDS_data_RT+HT_cooling}f).

To interpret these observations, the Schnakenberg model for a conduction process due to polaronic defect hopping \cite{Schnakenberg.1968} is considered again. Within this framework, there are three regions to be expected for the behavior of $\sigma\mtfg{dc}$ with temperature: at high temperatures, $\sigma\mtfg{dc} \propto \exp\left(-\frac{W\mtfg{H}}{k\mtfg{B}T}\right)$, where $W\mtfg{H}$ is the average energy barrier between two neighboring defect sites. Due to spatial overlap of the polaronic wave functions, $W\mtfg{H}$ is dependent upon the average distance between two sites, decreasing with rising defect density. Hopping is achieved by multi-phonon absorption (acoustical and optical) in this range. As the temperature drops to around $\frac{\theta\mtfg{D}}{2}$, the number of available optical phonons is lowered such that the critical hop is determined by single optic phonon absorption; here, the conductivity is proportional to the number of available optic phonons, $\sigma\mtfg{dc} \propto \exp\left(-\frac{\hbar\omega\mtfg{opt}}{k\mtfg{B}T}\right)$, with $\omega\mtfg{opt}$ a typical optical phonon frequency. As the temperature decreases further, only acoustic phonons are available; the activation energy for hopping in this range is of the order of the energy width $\Delta\epsilon$ of the acceptor distribution. This behavior is most clearly shown by the low-disorder sample: here, above approximately \SI{250}{\kelvin}, $\sigma\mtfg{dc}$ is activated with $E\mtfg{a\sigma,h} = \SI{378}{\milli\electronvolt}$, while at low temperature, the conductivity shows a very weak temperature dependence with $E\mtfg{a\sigma,l} \approx \SI{2.1}{\milli\electronvolt}$. These two regions are separated by a temperature of approximately \SI{200}{\kelvin}, which is indeed about half the Debye temperature. This temperature was already identified as a critical one during the discussion of the DC electric properties above. A similar change in activation energy can be seen for the RT-grown NiO, although it is not as pronounced. This can be explained by the much broader distribution of acceptor states in this sample due to a higher degree of disorder. For the HT-grown NiO, the distribution of acceptor states is highly confined in energy. For this reason, the VRH model cannot be applied to this sample, as was done before for the RT-grown one, because for this model the defect states have to be randomly distributed both spatially and energetically.

In the temperature range around $\frac{\theta\mtfg{D}}{2}$, the conductivity should vary as $\exp\left(-\frac{\hbar\omega\mtfg{opt}}{k\mtfg{B}T}\right)$. This is difficult to evaluate because the region between the high- and low-temperature activated terms is narrow. However, the occurrence of the second relaxation process in the low-disorder sample can be taken as an indication for single-phonon processes determining the critical hop because the activation energy of the relaxation process is close to the energies of the optical phonons at the $\Gamma$ point of NiO: $\hbar \omega\mtfg{TO} \approx \SI{49}{\milli\electronvolt}$, $\hbar \omega\mtfg{LO} \approx \SI{72}{\milli\electronvolt}$ \cite{Gielisse.1965}. Extrapolating the relaxation rates to $T\mhog{-1} = 0$ gives an attack frequency of the hopping process of ${\omega\mtfg{\infty} = \SI{4.23\pm 0.96 e7}{\radian\per\second}}$. According to the standard small-polaron theory \cite{Long.1982}, the temperature dependence of the maximum of dielectric loss due to hopping is

\begin{align}
\omega\mtfg{rel} &= \omega\mtfg{\infty}\exp\left(-\frac{E\mtfg{a}}{k\mtfg{B}T}\right) \nonumber \\
&= \omega\mtfg{opt}\exp\left(-2\alpha R\right)\exp\left(-\frac{E\mtfg{a}}{k\mtfg{B}T}\right)
\label{eq:SPH_peak_frequencies}
\end{align}

where the first exponential term represents the magnitude of spatial overlap between two sites at a distance $R$. Using Eqn.~(\ref{eq:SPH_peak_frequencies}) and estimating $R = \left(\frac{3}{4 \pi} N\mtfg{net}\mhog{-1}\right)\mhog{\frac{1}{3}} = \SI{6.2}{\nano\meter}$ for the HT-grown NiO film, the localization length $\alpha\mhog{-1}$ of the charge carriers localized at one V\tfg{Ni} acceptor site can be estimated as $\alpha\mhog{-1} \approx \SI{0.85\pm 0.01}{\nano\meter} \approx 2a_0$. This value is close to the one obtained from the DC measurements on RT-NiO in the first section of this paper where \SI{0.66}{\nano\meter} were obtained. The charge carriers in the RT-grown film seem to be more localized than in the low-disorder sample.

Lunkenheimer \textit{et al.} have investigated the AC electrical conductivity of reactively e-beam-evaporated NiO thin films and have applied the model of correlated barrier hopping (CBH) to explain the observed frequency and temperature dependence of $\sigma\mhog{*}$ \cite{Lunkenheimer.1991}. The authors also reported on a relaxation process visible at audio frequencies with a relaxation rate that is activated with an energy of around \SI{300}{\milli\electronvolt}. They interpreted this as a manifestation of the cut-off of the carrier hopping process. From the data given in the article, however, it can be estimated that the relaxation strength of the process is $\Delta\varepsilon \approx 5$ which is too large for a molecular process. One may therefore discuss whether Lunkenheimer \textit{et al.} also measured the dielectric response of a system exhibiting a spatially inhomogeneous conductivity. In any case, it is interesting to note the physical similarity of the CBH model and the defect hopping model. Both consider the transfer of charge carriers between neighboring acceptor (or donor) sites by thermal activation over an energy barrier the height of which depends on inter-site separation.\\

Two comments on the DC conductivities $\sigma\mtfg{dc}$ measured by broadband dielectric spectroscopy (BDS) shall be made. The first one concerns the agreement between these results and those obtained from $IV$ measurements in the first section. They agree very well in the temperature range between \SI{300}{\kelvin} and \SI{200}{\kelvin}. Below that temperature, however, $\sigma\mtfg{dc}$ as extracted from BDS measurements decrease more strongly than in the data obtained from $IV$ characteristics; at $\SI{120}{\kelvin}$, the values differ by about one order of magnitude. This is due to the fact that $IV$ characteristics are not measured under static conditions, because a voltage ramp is swept through. Therefore, frequency-dependent currents, like the MWS polarization process, can contribute to the current signal. If the $\log \sigma\mtfg{dc}$ values from BDS measurements are plotted versus $T\mhog{-\frac{1}{4}}$, however, the $T_0$ value characterizing the VRH conduction, and thereby also the localization length $\alpha\mhog{-1}$, are very similar to the one obtained before ($T_0 = \SI{3.63\pm 0.04 e9}{\kelvin}$ and $\alpha\mhog{-1} = \SI{0.63\pm0.02}{\nano\meter}$, respectively) such that the main assertions of the first section remain essentially untouched.

The second comment concerns the difference of $\sigma\mtfg{dc}$ between the HT- and the RT-grown samples at low temperatures. One may wonder why the former one exhibits a higher conductivity in this range, although the density of acceptor states is lower by a factor of about four. The answer is provided by the width of the energetic distribution of the impurities, which is much smaller in the weakly compensated HT-grown film ($\Delta\epsilon\mtfg{HT-PLD} \approx \SI{2}{\milli\electronvolt}$ vs. $\Delta\epsilon\mtfg{RT-NiO} \approx \SI{230}{\milli\electronvolt}$). The better energetic alignment of the states facilitates a direct transfer of carriers between neighboring impurities by tunneling. This change in transport character from hopping to tunneling is also reflected by the fact that the position of both dielectric relaxation processes is not coupled to the magnitude of $\sigma\mtfg{dc}$ in the low-temperature regime: both processes continue to shift to lower frequencies until they are partially or even completely masked by the constant-conductivity term. 

\begin{figure}
	\centering
	\includegraphics[width=\columnwidth]{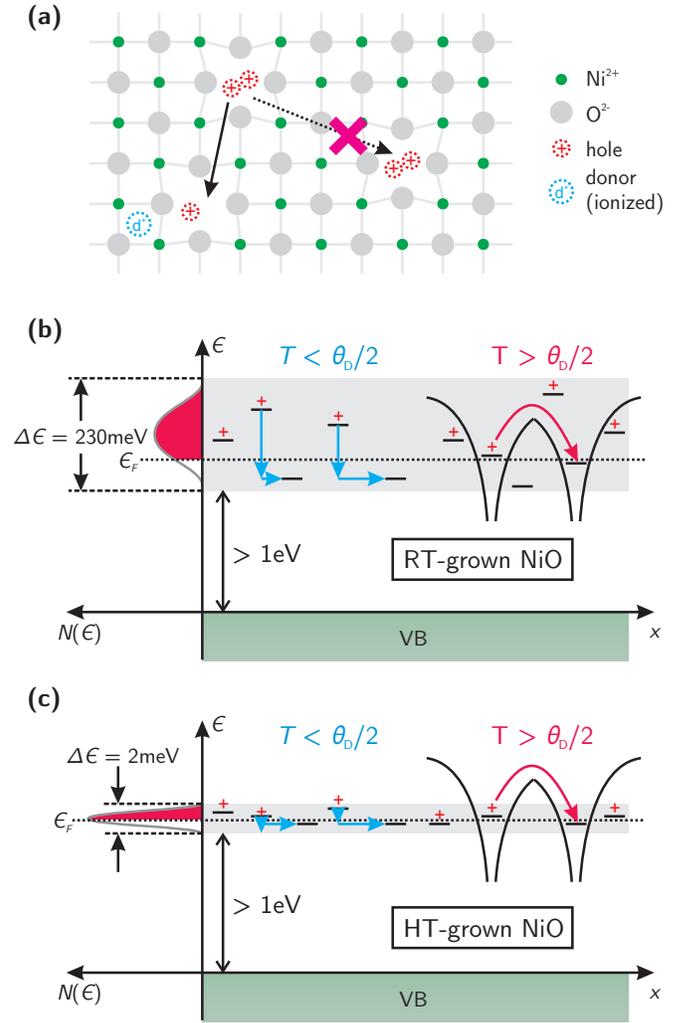}
	\caption{Illustration of the proposed conduction model. (a) Transport is carried by holes propagating exclusively on Ni vacancy sites. Partial compensation by donors is necessary to enable charge transfer. (b) High degree of structural and electronic disorder introduces a broad distribution of deep lying defect states in RT-grown NiO. For the low-disorder sample (c), the acceptor states are considerably more confined in energy. The temperature $T = \frac{\theta\mtfg{D}}{2} \approx \SI{200}{\kelvin}$ separates two regimes with different hopping characteristics. Energies and spatial distances are not to scale.}
	\label{fig:scheme}
\end{figure}

\subsection{Summary of the proposed transport model}

The conduction model for intrinsically doped NiO constructed in this paper is summarized in Fig.~\ref{fig:scheme}. It is based on holes trapped on V\tfg{Ni} sites due to the formation of (small-polaron like) bound Zhang-Rice states, and their direct transfer between such sites (Fig.~\ref{fig:scheme}a). Because of the strongly bound character of these quasi-particles, any extended valence band states are sufficiently far away in energy that they can be neglected in the conduction process. At high temperature ($T > \frac{\theta\mtfg{D}}{2}$), the charge carriers hop over the energetic barrier between defects by multi-phonon absorption. The activation energy of this process is similar for high- and low-disorder NiO because it depends only on the inter-site separation. When $T<\frac{\theta\mtfg{D}}{2}$, hopping over the barrier is suppressed; in this regime, the current is carried by phonon-assisted tunneling through the barrier. The activation energy for this process is given by the energetic broadening of the defect state distribution which is lower for the low-disorder sample (Fig.~\ref{fig:scheme}c). Because most states lie within a narrow interval around the Fermi energy, the current carried by this process is more efficient than in the high-disorder, room-temperature-grown sample where fewer defect states with a similar energy level are available. Furthermore, for this sample, the states are widely distributed such that the VRH model applies at all temperatures below \SI{300}{\kelvin}. In contrast to this, for the HT-NiO film, the narrow distribution of states allows charge transfer only between nearest neighbor defects.

Based on this model, an explanation can be given why there are some reports on NiO in the literature which claim hole mobilities of more than \SI{1}{\square\centi\meter\per\volt\per\second}, far outside the range typical for polaronic hopping transport \cite{Sato.1993,Molaei.2013,Chen.2015,Tyagi.2015}. In many of these works, the conductivity of the investigated films is in the range of \SI{1e-2}{\siemens\per\centi\meter} and higher, suggesting higher doping levels and thereby a stronger spatial overlap of the ZR wave functions than reported in this work. Also, sputtering at room temperature is typically used, leading to films with high structural and electronic disorder. The result is a broad band of V\tfg{Ni} defect states. The main contribution of the conductivity can still be expected to come from hopping processes; however, a small part of the carriers occupy levels lying close enough in energy around $\epsilon\mtfg{F}$ that they can move in a very narrow band by almost activationless transport, similar to what was observed in our HT-NiO sample at low temperatures (Fig.~\ref{fig:sigmaDC+BNN}a). These carriers are able to generate a Hall voltage, while the hopping carriers are not. The hole densities extracted from these measurements are in the range of \SI{e19}{\per\cubic\centi\meter} or larger, but for the reason just explained, this is only a small subset of the overall hole population. Because the conductivity, on the other hand, is determined by both carrier types and therefore rather large, the calculated "Hall mobility" $\mu\mtfg{Hall} = \frac{\sigma}{ep\mtfg{Hall}}$ overestimates the value attributed to the hopping-determined conductivity.

	\section{Conclusion}
	
	The temperature dependence of the DC and AC conductivity of intrinsically doped thin films can be consistently explained within the framework of polaronic inter-acceptor hopping conduction. This model is motivated by the perception of the nature of holes in doped NiO, which has been shown to be of strongly bound character (Zhang-Rice bound doublet). The ZR states are therefore situated deep inside the band gap, sufficiently well isolated from any extended valence band such that charge transfer can only occur directly between the acceptors. The degree of structural and electronic disorder is of great importance for the conduction process. For samples of high disorder, Mott variable-range hopping (VRH) can be applied well to model the behavior of $\sigma\mtfg{dc}$ due to a large spread of the acceptor energies. VRH can be seen as a limiting case of the inter-acceptor hopping model for a broad distribution of acceptor states. In the case of low disorder, there is a clear change of the temperature behavior of $\sigma\mtfg{dc}$ at about half of the Debye temperature, $\frac{\theta\mtfg{D}}{2} \approx \SI{200}{\kelvin}$, with multi-phonon thermal activation above and phonon-assisted tunneling below that temperature, in accordance with the predictions of polaronic inter-acceptor hopping. Additionally, around and below this temperature the low-disorder sample exhibits a dielectric relaxation process in $\varepsilon''$ which is connected to the frequency cut-off of single-phonon-assisted hopping. The results from the VRH model and the parameters of this process give a localization length of the ZR holes of \SI{0.62}{\nano\meter} and \SI{0.85}{\nano\meter}, or about two lattice constants, for RT- and HT-grown NiO, respectively. 
	
	Furthermore, it is shown that disorder leads to the occurrence of strong interfacial polarization of mesoscopic regions in the film due to inhomogeneous conductivity (Maxwell-Wagner-Sillars polarization). This effect can produce a high apparent dielectric response under low-frequency AC electric fields.
	The current-voltage characteristics of NiO films sandwiched between Pt electrodes are non-ohmic but symmetrical, with an increase of conductivity at higher DC electric fields which is observed at all temperatures. This could be attributed to a linear increase of the density of ZR states by hole injection. In these measurements, the temperature $\frac{\theta\mtfg{D}}{2}$ seems to be equally critical as in the AC measurements, because it marks the onset of space-charge-limited conduction due to increasing carrier trapping time.

	\begin{acknowledgments}
	This work was funded by the Deutsche Forschungsgemeinschaft (DFG) in the framework of the collaborative research centers "SFB 762: Functionality of Oxidic Interfaces" (project B06) and "SFB/TRR 102: Polymers under multiple constraints" (project B08).
	\end{acknowledgments}
	
	\bibliography{ref}
	
\end{document}